# Smart Contract Vulnerability Detection based on Static Analysis and Multi-Objective Search

Dongcheng Li
Department of Computer Science
California State Polytechnic
University - Humboldt,
Arcata, USA

W. Eric Wong*
Department of Computer Science
University of Texas at Dallas
Richardson, USA

Xiaodan Wang
School of Computer Science
China University of Geosciences
Wuhan, China

Sean Pan and Liang-Seng Koh
RFCyber Corporation
Frisco, USA

*Abstract*—This paper introduces a method for detecting vulnerabilities in smart contracts using static analysis and a multi-objective optimization algorithm. We focus on four types of vulnerabilities: reentrancy, call stack overflow, integer overflow, and timestamp dependencies. Initially, smart contracts are compiled into an abstract syntax tree to analyze relationships between contracts and functions, including calls, inheritance, and data flow. These analyses are transformed into static evaluations and intermediate representations that reveal internal relations. Based on these representations, we examine contract's functions, variables, and data dependencies to detect the specified vulnerabilities. To enhance detection accuracy and coverage, we apply a multi-objective optimization algorithm to the static analysis process. This involves assigning initial numeric values to input data and monitoring changes in statement coverage and detection accuracy. Using coverage and accuracy as fitness values, we calculate Pareto front and crowding distance values to select the best individuals for the new parent population, iterating until optimization criteria are met. We validate our approach using an open-source dataset collected from Etherscan, containing 6,693 smart contracts. Experimental results show that our method outperforms state-of-the-art tools in terms of coverage, accuracy, efficiency, and effectiveness in detecting the targeted vulnerabilities.

## I. INTRODUCTION

With the rapid expansion of smart contract applications and their direct economic implications, the frequency of attacks on smart contracts has increased significantly. As a nascent software paradigm, smart contracts are prone to vulnerabilities [1]. Unlike traditional program code, which can be deleted or modified, blockchain-based smart contracts are immutable once deployed, making it impossible to fix vulnerabilities in already deployed contracts [2]. This necessitates thorough testing of smart contracts before deployment to enhance their reliability [3].

To address vulnerabilities in smart contracts, researchers have focused on two primary areas. First, various methods have been proposed to identify errors in smart contracts, including formal verification, symbolic execution, fuzz testing, and deep learning [4]. Second, techniques for repairing detected vulnerabilities based on identified error types have gradually developed [5]. Despite these efforts, many existing detection tools and repair methods struggle with automation and scalability, leaving significant threats unaddressed [6].

Building on this foundation, the primary challenges in smart contract vulnerability detection include the complexity of smart contract code, ensuring the accuracy of detection results, the limited availability of high-quality annotated datasets, and the resource-intensive nature of the detection process. Analyzing complex code with varied functions, intricate transaction processes, and diverse data types and control structures is difficult and time-consuming. Ensuring accurate detection results is crucial to avoid false positives or negatives, which can lead to significant security risks. Additionally, the detection process requires substantial time and computational resources, often necessitating professional review, making it less practical for large-scale or real-time applications.

In light of this background, this paper proposes a method for detecting smart contract vulnerabilities based on static analysis and multi-objective optimization algorithms. This approach involves analyzing the characteristics of smart contracts and examining the dependencies and data flows between contracts and functions. By employing multi-objective optimization algorithms, we track the data flow and value changes of different input data to monitor code coverage and the number of newly discovered code blocks during the analysis. The accuracy of detected vulnerabilities is calculated under different individual environments. The results are validated using datasets collected from existing literature and Etherscan [7], aiding software developers and testers in applying targeted repair methods and improving the efficiency of software maintenance. More specifically, this paper makes the following contributions:

- We use static analysis methods to detect vulnerabilities in smart contracts. By converting Solidity source code into an abstract syntax tree, we perform internal analysis, evaluation, and intermediate representation to identify vulnerabilities such as reentrancy, call stack overflow, integer overflow, and timestamp dependency.

- We propose a vulnerability detection technique that combines static analysis with a multi-objective optimization algorithm. This approach monitors data flow changes during execution and applies operations such as crossover, mutation, and selection to the input data population. The algorithm improves statement coverage and detection accuracy during the vulnerability detection process.

- We compile a dataset of annotated smart contracts to validate our method. The proposed method is compared with state-of-the-art detection tools like Mythril and

Oyente, demonstrating superior performance in terms of accuracy, detection efficiency, and overall effectiveness.

The structure of the rest of the paper is organized as follows: Section II reviews the current state of research on smart contract vulnerability detection. Section III discusses the static analysis-based detection framework, explaining the multi-objective optimization algorithm's applicability in the context of static analysis for vulnerability detection. Section IV presents the experimental setup, datasets, detection results, and analysis of the performance of the proposed method. Section V concludes the paper by summarizing the findings, addressing research limitations, and offering future research directions.

## II. RELATED STUDY

Numerous scholars have adopted various methods to analyze and identify vulnerabilities in smart contracts. Among them, program analysis techniques detect vulnerabilities by analyzing the smart contract itself, which can be divided into dynamic and static analysis. Dynamic analysis evaluates whether a program has issues based on various runtime states, while static analysis assesses potential problems by examining the code's control flow and data flow processes [8].

SmartCheck [9] is a static analysis tool designed to identify vulnerability patterns and poor coding practices in Ethereum smart contracts written in Solidity. The tool first converts Solidity source code into XML (Extensible Markup Language) for intermediate representation, then checks it using XPath patterns. The authors conducted experiments using validated large datasets and compared the results with three different manual audits. However, SmartCheck's detection capabilities are limited, often requiring supplementary techniques and manual intervention, which reduces its level of automation.

Slither [10] is a static analysis framework that converts Solidity smart contracts into an intermediate representation called SlithIR. It applies various program analysis techniques, such as data flow and taint tracking, to extract and refine information. Slither aids in code reviews, enhances user comprehension, and automatically detects potential code optimizations and vulnerabilities.

Vandal [11] is a security analysis tool for detecting vulnerabilities in smart contract bytecode. It features an analysis pipeline that converts low-level Ethereum Virtual Machine (EVM) bytecode into higher-level logical relations. This process includes scraping bytecode from the blockchain, disassembling it into opcodes, decompiling it into a register transfer language, and extracting logic relations that capture the program's semantics. By leveraging the Soufflé language, Vandal enables declarative security analyses, which enhances its speed and robustness.

SDCF (Software-Defined Cyber Foraging) [12] is a framework that integrates static and dynamic analysis techniques with taint tracking to provide efficient runtime protection and achieve high code coverage while detecting potential software vulnerabilities. Despite its effectiveness, SDCF cannot handle context-dependent control transfers, which limits it from achieving higher code coverage.

Program analysis techniques, similar to feature-based code matching [13], offer fast detection but insufficiently extract semantic information from smart contract code, resulting in lower accuracy. Moreover, smart contracts exhibit more functional and feature variations compared to traditional program code, making pure analysis less effective.

## III. SMART CONTRACT VULNERABILITY DETECTION BASED ON MULTI-OBJECTIVE SEARCH AND STATIC ANALYSIS

Due to the unique characteristics of smart contracts, such as their immutability post-deployment and significant security risks, existing vulnerability detection tools for traditional programming languages cannot be directly applied to them.

### A. Static Analysis Framework for Smart Contract Vulnerability Detection

Static analysis involves analyzing the source code of smart contracts to detect potential vulnerabilities. This study uses the Solidity Abstract Syntax Tree (AST) generated from the smart contract source code as the initial input for the static analyzer [14]. Compared to the source code, the AST provides a more intuitive representation of the code's syntax structure with fixed node types. Additionally, since Solidity is a strongly typed language with strict and standardized syntax rules, transforming the program into an AST better preserves the context and semantic information of the source code.

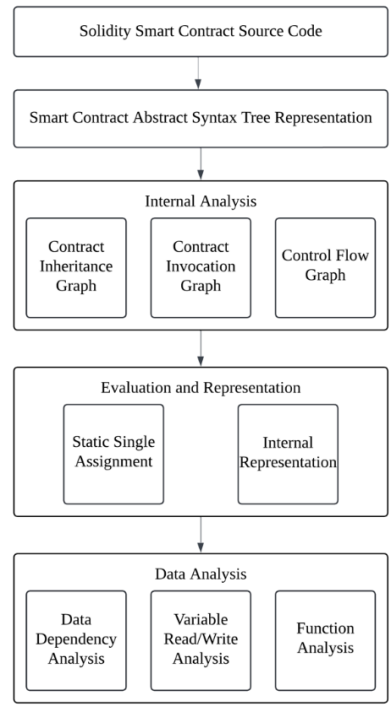

Fig. 1. Static analysis framework for smart contract vulnerability detection

The framework for static analysis-based smart contract vulnerability detection is shown in Fig. 1 and includes three main modules: the Internal Analysis Module, the Evaluation and Representation Module, and the Data Analysis Module.

- Internal Analysis Module: This module analyzes the internal structure of the contract, examining the inheritance and call relationships between subcontracts based on the contract's AST and calculating the storage order of data variables [15,16].

- Evaluation and Representation Module: This module uses Static Single Assignment (SSA) to facilitate computational analysis of the code and represents the code structure using intermediate representation language [10,17].

- Data Analysis Module: This module provides enhanced information for each module through predefined function analysis, using methods such as data dependency and variable analysis to further analyze the contract and identify vulnerability types [18-27].

*B. Issues in Static Analysis for Smart Contract Vulnerability Detection*

Compared to dynamic analysis, static analysis methods for smart contract vulnerability detection offering broader coverage of the code and the ability to detect vulnerabilities that are difficult to identify through dynamic analysis. Static analysis can detect vulnerabilities before the code is executed, allowing for early identification and fixing of potential vulnerabilities before the code goes live, thereby avoiding potential security risks. However, the current static analysis methods for smart contract vulnerability detection still have the following issues:

- Static analysis methods may produce false positives and false negatives because smart contracts are uploaded to the blockchain in the form of EVM bytecode, which is a stack-based low-level code characterized by dynamic code creation and invocation. It is challenging to accurately determine whether certain codes have vulnerabilities without knowing the execution environment and input data of the program.

- Static analysis methods can extract the syntactic and semantic information of smart contract codes, but they lack dynamic data flow analysis, leading to the omission of some branch jump statements, thus it has the issue of Low Statement Coverage.

- Smart contracts use complex language and has complex logic, where a single data analysis method may only be suitable for detecting one type of vulnerability. Different range analysis methods show significant differences in performance in detecting various types of vulnerabilities.

*C. Applicability of Multi-Objective Search in Static Analysis for Vulnerability Detection*

The static analysis for smart contract vulnerability detection can be transformed into a multi-objective optimization problem [28]. The NSGA-II algorithm, a fast non-dominated sorting genetic algorithm with an elitist strategy, is a multi-objective optimization algorithm based on Pareto optimal solutions. Since the aim of this study is to simultaneously optimize two objectives, the NSGA-II algorithm is chosen for this study.

*1) Chromosome Encoding Design*

The key to transforming the smart contract vulnerability detection problem into a multi-objective optimization problem lies in chromosome encoding design. This study optimizes vulnerability detection capability through the original input data of smart contracts and adopting real number encoding.

When designing chromosome encoding, each gene represents an original input data of the smart contract. A chromosome is all original input data in a smart contract. The variables are stored sequentially and then converted into a chromosome, as shown in Fig. 2.

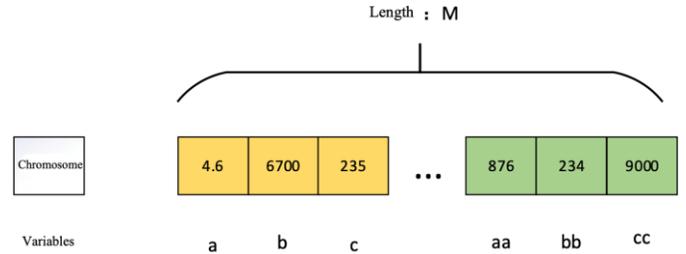

Fig. 2. Chromosome encoding

The chromosome length M is the number of original input data of the current smart contract being detected. During the static analysis stage, the initial original data involved in each function and contract of the smart contract program are recorded. During encoding, one data is filled into one gene. The significance of an individual population is the original input and boundary values of all data involved in the smart contract program, with data changes restricted by boundary values.

*2) Constraints and Objective Functions*

*a) Constraints*

During algorithm operation, the changes of chromosomes are restricted by a series of constraints, including equality and inequality constraints. The inheritance and invocation relationships of functions derived from the internal structure analysis of smart contracts will affect the data flow and statement coverage. Based on the specific analysis of smart contracts, the following three types of constraints are obtained:

- Boundary Values: The data changes must be confined to specified ranges according to their types. If the data exceed the specified range during changes, they need to be discarded or regenerated.

- Data Dependencies: During the analysis of data dependencies by static analysis methods, the contextual data dependencies of each function are stored in a graph. Changes in the original input data will affect other variables, necessitating the removal of data that do not satisfy dependency relationships.

- Inheritance Relationships: The inheritance relationships of contracts will affect the data generation conditions of base and derived classes. During data changes, the construction of derived classes is influenced by the base class. If derived classes lack specified parameters, the attributes of the base class determine the data of derived classes. Chromosomes that exceed the constraints of derived classes need to be excluded.

*b) Objective Functions*

The introduction of a multi-objective optimization algorithm aims to guide the path of original data changes, improving the statement coverage involved in vulnerability detection and enhancing detection accuracy during the process. Therefore, how to mutate and extend to more superior individuals and how to select superior individuals in non-dominated sorting are crucial for the NSGA-II algorithm to improve the static analysis method for vulnerability detection. The fitness functions are set to the accuracy of vulnerability detection and the code coverage. Static analysis might miss vulnerabilities due to the lack of dynamic execution; hence, expanding code coverage can help find potentially missed defects. However, focusing too much on statement coverage can lead to false positives, so balancing accuracy and coverage is crucial.

- *Accuracy*

$$max: f_1(x) = \frac{N}{\sum_{i=1}^{m} D_i} \quad (1)$$

$N$ represents the total number of detected vulnerabilities, $m=4$ refers to the four types of vulnerabilities, and $D_i$ denotes the number of one type of vulnerability. Accuracy is the total number of detected vulnerabilities divided by the total number of labeled vulnerabilities in the dataset

- *Statement Coverage*

$$max: f_2(x) = \frac{j+k}{\sum_{i=1}^{s} J_i + K} \quad (2)$$

$j$ denotes the number of all covered jump statements, $k$ is the number of covered regular statements, $s=6$ is the total number of jump statements, $J_i$ represents the number of a specific jump statement, and $K$ is the total number of regular statements.

*D. Multi-Objective Search for Static Analysis in Vulnerability Detection*

*1) Framework*

Based on the static analysis framework for smart contracts vulnerability detection, this study incorporates modules for the multi-objective search and data execution for static analysis in vulnerability detection, as shown in Fig. 3. Based on this, initial data such as original input, boundary values, and special data are generated, forming the initial population randomly.

The algorithm module optimizes the data obtained from static analysis using a multi-objective optimization algorithm, namely NSGA-II. It encodes the original input data of the smart contract with real numbers, and extends more potentially superior individuals through mutation and crossover. Based on this, individuals are ranked using non-dominated sorting, and the diversity of solutions is maintained through crowding distance calculation, selecting superior individuals. The ultimate goal is to cover more code blocks and detect previously uncovered vulnerabilities, improving detection accuracy and reducing false negatives in the static analysis framework.

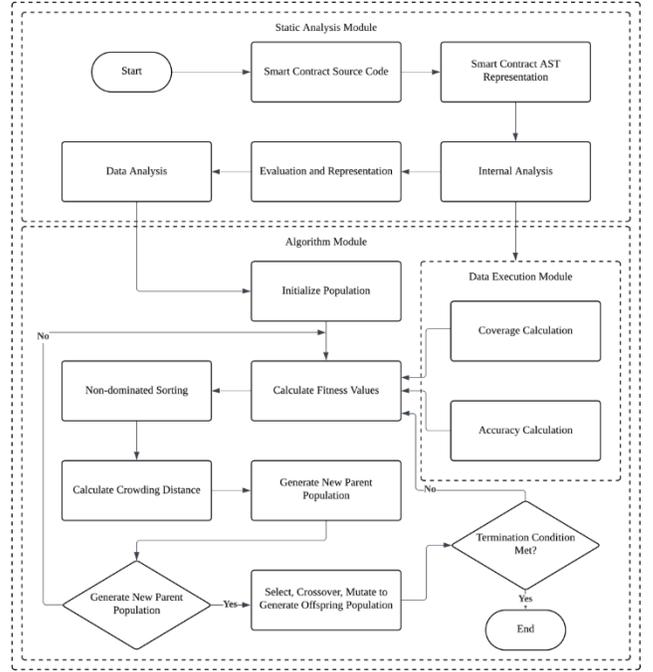

Fig. 3. Framework of multi-objective search for static analysis in vulnerability detection

The data execution module calculates the objective functions in the multi-objective optimization algorithm, which are coverage and accuracy. Coverage is determined by simulating the data flow execution process based on the dependency relationships in the static analysis module's data flow graph. The process begins at the data flow graph's entry, deducing all variable values and states step-by-step according to dependencies until the data flow graph's exit. During the simulation of data flow execution, the variable values and states are detected and verified to ensure the program's correctness and stability. Finally, the changes on execution path and variable value are extracted from the simulated data flow execution results, calculating the branch coverage and vulnerability detection accuracy.

*2) Algorithm Module*

*a) Encoding Method*

Since the algorithm optimizes vulnerability detection through the original input data of smart contracts, thus we adopt real number encoding. Each original input data of the smart contract corresponds to one gene. The chromosome length varies according to the number of original input data of the current smart contract being detected, matching it. The significance of a population individual is the original input and boundary values of all data involved in the smart contract program, with data changes restricted by boundary values.

*b) Population Initialization*

After obtaining the data dependency graph from the static analysis module's analysis of smart contracts, all initial input data in the graph are selected as genes to form chromosomes, generating a batch of populations according to the population size. The random initialization method, which is simple to implement and approximately uniform in population individuals, is adopted. The initialized population will gradually find superior individuals through the NSGA-II algorithm, preparing for the final optimization.

*c) Selection Operation*

The Pareto-domination-based selection in NSGA-II involves non-dominated sorting and crowding distance calculations. First, the fitness value, dominance set, dominance count, and crowding distance of each individual are calculated. The dominance set includes all individuals that the current individual dominates, with a smaller dominance set indicating a more competitive individual. The dominance count refers to the number of individuals that dominate the current individual. Crowding distance, the average distance between the current individual and its neighbors, is used to maintain population diversity. Next, non-dominated sorting is performed, dividing individuals into different levels (Pareto fronts), each containing non-dominated individuals. Within each Pareto front, crowding distance is calculated to preserve diversity. Finally, individuals are selected based on Pareto front and crowding distance, choosing those with higher-ranking Pareto fronts and larger crowding distances. This process repeats until the required number of individuals is selected, forming the next generation population.

*d) Crossover Operation*

During crossover operation, a uniform crossover method is used. For each gene, a certain probability is used to choose whether to inherit the gene from one parent or the other. Two new individuals obtained at the end of the crossover are returned as the new population.

*e) Mutation Operation*

The new gene values generated by mutation may exhibit superior results in multiple objective functions. The population queue that reaches this step generates mutation values through uniform integer mutation [29] and random order mutation [30]. Uniform integer mutation involves making random changes with uniform distribution to each gene in an individual. A uniformly distributed random number is generated within the range and added to the original gene value to obtain the mutated new gene value. If the new gene value exceeds the range, it is truncated. The new individual obtained after mutation is returned as the next generation population. The mutation rate of uniform integer mutation is fixed and cannot adapt adaptively, which may result in excessively large or small mutation amplitudes, thereby affecting the search performance.

Random order mutation involves randomly selecting two positions within the chromosome and swapping their corresponding segments to create a new individual. The selected segments are then reordered randomly, forming a new sequence. This new sequence replaces the original segments, resulting in a new individual. The length and selection criteria of the segments can be adjusted during experiments to enhance algorithm efficiency and search performance.

*3) Data Execution Module*

The data execution module simulates data execution to calculate code coverage and final vulnerability detection accuracy. This module derives from the data dependency relationships obtained from static analysis, including the definition and usage of variables in the smart contract program, and the data transmission and dependencies between variables. Code coverage in this study refers to statement coverage, where a statement is a complete operation, possibly including multiple basic blocks and jump instructions. Jump instructions in smart contracts implement the program's control flow, with primary instructions listed in Table I.

TABLE I. JUMP INSTRUCTIONS

| No. | Jump Instruction | Description |
|---|---|---|
| 1 | JUMP | Unconditional jump to a specified position for execution, typically used for loops and conditionals. |
| 2 | JUMPI | Conditional jump to a specified position if a condition is met, used for implementing conditional branches. |
| 3 | JUMPDEST | Marks a valid target for JUMP and JUMPI instructions, ensuring a safe jump destination. |
| 4 | RETURN | Ends execution and returns data to the caller, commonly used to return output from functions. |
| 5 | REVERT | Ends execution, reverts state changes, and optionally returns an error message, used for error handling. |
| 6 | STOP | Halts execution without reverting state changes, used to end contract execution safely. |

Jump instructions are widely used in smart contracts, enabling various complex control flows, such as conditional statements, loop statements, exception handling, function calls, etc. Therefore, jump instructions' usage needs comprehensive coverage for smart contract analysis and testing.

This study employs an instrumentation method to calculate smart contract statement coverage. Different calculation methods are used for regular and jump statements, with the algorithmic process as follows.

| Instrumentation Algorithm Pseudo-Code |
|---|
| *Input*: Smart contract |
| *Output*: Statement coverage |
| 1:    Analyze the source code of the smart contract and generate the control flow graph of basic blocks. |
| 2:    Insert counter code at the entrance of each basic block to record the execution times of each basic block. |
| 3:    Insert counter code before and after each statement within basic blocks to record the execution times of each statement. |
| 4:    *For each* statement: |
| 5:       *if* statement is a jump instruction: |
| 6:          Insert counters before and after the jump instruction's target position |
| 7:       *else* |
| 8:          Insert counters for regular statements |
| 9:       *End* |
| 10:    *End* |
| 11:   Run the data, and statistically analyze the counters to calculate the coverage of each basic block and statement. |
| 12:   Calculate the statement coverage based on the counters of all statements and covered statements. |
| 13:   Return statement coverage. |

Calculating the accuracy of smart contract vulnerability detection requires high-quality datasets containing labeled vulnerabilities. To obtain such high-quality dataset, we build a labeled dataset that incorporates various publicly available labeled datasets sourced from the literature. Using such datasets, the detected vulnerabilities are compared with labeled data, and statistical analysis is performed to obtain accuracy. Accuracy is simply calculated by dividing the number of correctly detected vulnerabilities by the total number of vulnerabilities. Since smart contract vulnerability detection has certain false positive and false negative rates, only the accuracy of vulnerability detection is considered in the multi-objective optimization objective function section.

## IV. EXPERIMENT RESULT AND ANALYSIS

To demonstrate the effectiveness of incorporating a multi-objective optimization algorithm into the static analysis method for vulnerability detection, various experiments were conducted. These experiments evaluate the performance improvements, efficiency, and the method's applicability on unlabeled datasets.

### 1) Experiment Design

#### a) Research Questions

This study aims to verify the following research questions:

- RQ1: Does the NSGA-II algorithm improve the coverage and accuracy of vulnerability detection in static analysis and to what extent?
- RQ2: How does the proposed method perform in comparison to other tools in terms of vulnerability detection efficiency?
- RQ3: How does the proposed method compare to other similar methods used for vulnerability detection in terms of overall performance?
- RQ4: How effectively and accurately does the proposed method identify vulnerabilities in unlabeled smart contracts across different criteria?

#### b) Datasets

It is evident from the existing literature that the number of publicly available smart contracts with labeled vulnerabilities is limited [22,31-34]. Datasets labeled using detection tools may not always be reliable, and manual labeling is labor-intensive, resulting in a limited number of labeled contracts. Therefore, this study utilizes two types of datasets: manually labeled datasets and unlabeled smart contracts, as mentioned in previous sections.

The source code of manually labeled smart contracts primarily comes from Chen et al. [22], where the authors manually identified defects with its types and numbers in 587 real-world smart contracts and made it publicly available. Zhang et al. [31] manually wrote 176 smart contracts, including faulty contracts, fixed contracts, and carefully crafted contracts that appear vulnerable but are not. Durieux et al. [32] collected 69 contracts from GitHub, blog posts, and Etherscan, labeled 115 vulnerabilities, and categorized them into ten types.

After filtering for similarity and excluding contracts that were problematic or non-executable, we finalized a dataset of 623 viable contracts. These contracts formed the basis for our experiments on manually labeled datasets.

The unlabeled smart contracts were sourced from real smart contracts on the Ethereum network from Etherscan. Ferreira et al. [33] retrieved the source code of smart contracts associated with addresses through Ethereum's transaction data and API, resulting in 47,518 unique smart contracts after removing duplicates, including 623 manually labeled contracts. Huang et al. [34] downloaded 32,537 Solidity files from Etherscan and calculated their syntactic and semantic similarities to identify similar smart contracts, eventually matching code changes in 42 versions of evolving smart contracts. After preprocessing and excluding the labeled contracts, we selected 6,070 Solidity files for experiments.

In total, 6,693 smart contracts were compiled, including 623 manually labeled contracts and 6,070 unlabeled contracts. The line count of all Solidity files was calculated, classifying contracts with fewer than 50 lines as simple contracts, those with 50-300 lines as ordinary contracts, and those with more than 300 lines as complex contracts. The dataset statistics used in the experiments are shown in Table II.

TABLE II. SMART CONTRACT LINE COUNT STATISTICS

| Contract Type | Simple Contracts | Ordinary Contracts | Complex Contracts | Total |
|---|---|---|---|---|
| Manually Labeled | 66 | 251 | 306 | 623 |
| Unlabeled | 760 | 2917 | 2339 | 6070 |
| Total | 826 | 3168 | 2645 | 6693 |

#### c) Experimental Environment

The experiments were divided into two categories: one based on labeled data and the other on unlabeled data. After conducting multiple experiments, the parameters set for NSGA-II algorithm are presented in Table III. These parameter values are based on empirical data obtained from repeated experiments.

TABLE III. NSGA-II PARAMETER TABLE

| Parameter | Value | Description |
|---|---|---|
| $N$ | 50 | Population size |
| $T$ | 200 | Maximum iterations |
| $K$ | 20 | Number of selected individuals |
| $P_C$ | 0.6 | Crossover rate |
| $P_m$ | 0.75 | Mutation rate |

The specific hardware and runtime environment information for the experiments are shown in Table IV.

TABLE IV. EXPERIMENTAL ENVIRONMENT SETUP

| Environment | Parameters |
|---|---|
| Operating System | Ubuntu 20.04.2 |
| Python Version | 3.8 |
| Py-solc Version | 0.4.4 - 0.6.10 |

#### d) Evaluation Metrics

Vulnerability detection for a specific vulnerability type is essentially a binary classification problem—whether the vulnerability exists or not. For such problems, the match between the actual condition and the experimental detection results is an important evaluation metric. This is typically represented using a confusion matrix, as defined in Table V.

TABLE V. CONFUSION MATRIX

| Confusion Matrix | | Actual Results | |
|---|---|---|---|
| | | Positive | Negative |
| Detected Results | Positive | TP | FP |
| | Negative | FN | TN |

- TP (True Positive): Cases where detected vulnerabilities that actually exist.
- FP (False Positive): Cases where detected vulnerabilities that do not actually exist.
- FN (False Negative): Cases where no vulnerability was detected, but one exists.
- TN (True Negative): Cases where no vulnerability was detected, and none exists.

Based on the confusion matrix, the evaluation metrics used in this experiment include: Accuracy, Precision, Recall, F1-Score, ROC (Receiver Operating Characteristic Curve), AUC (Area Under Curve), MCC (Matthews Correlation Coefficient), and FMI (Fowlkes and Mallows Index). The specific formulas for each metric are as follows

**Accuracy**: The percentage of correct predictions out of the total number of samples.

$$accuracy = \frac{TP+TN}{TP+TN+FP+FN} \quad (3)$$

**Precision**: The proportion of positive predictions that are actually correct, representing the probability that a detected vulnerability is indeed a vulnerability.

$$precision = \frac{TP}{TP+FP} \quad (4)$$

**Recall**: The proportion of actual positive cases (vulnerabilities) that are correctly detected by the model.

$$recall = \frac{TP}{TP+FN} \quad (5)$$

**F1-Score**: A metric that balances precision and recall, calculated as the harmonic mean of precision and recall.

$$F1_{Score} = \frac{2*precision*recall}{precision+recall} \quad (6)$$

The AUC of the ROC curve is based on the True Positive Rate (TPR) and False Positive Rate (FPR). TPR represents the proportion of positive class samples that are correctly detected as positive, while FPR represents the proportion of negative class samples that are incorrectly detected as positive. FNR (False Negative Rate) and TNR (True Negative Rate) represent the proportion of positive samples incorrectly detected as negative and the proportion of negative samples correctly detected as negative, respectively. The formulas for TPR, FPR, FNR, and TNR are given in equations (7) to (10). The vertical axis of the ROC is the TPR, and the horizontal axis is the FPR. The AUC is the area under the ROC curve, normalized to 1, representing its value.

$$TPR = \frac{TP}{TP+FN} \quad (7)$$

$$FPR = \frac{FP}{FP+TN} \quad (8)$$

$$FNR = \frac{FN}{TP+FN} \quad (9)$$

$$TNR = \frac{TN}{FP+TN} \quad (10)$$

MCC is considered an unbiased version of the F1-Score, as it uses all elements of the confusion matrix. It is typically regarded as a balanced metric for binary classification, applicable even when the sample sizes of the two classes differ significantly. It essentially describes the correlation coefficient between actual and predicted classifications, ranging from -1 to 1, where 1 indicates perfect prediction, 0 indicates no better than random prediction, and -1 indicates complete inconsistency between predicted and actual classifications. The MCC formula is shown in equation (11).

$$MCC = \frac{TP*TN-FP*FN}{\sqrt{(TP+FP)(TP+FN)(TN+FP)(TN+FN)}} \quad (11)$$

FMI is an external index used to compare clustering results with a reference model, measuring the similarity between two hierarchical clusters or between clusters and benchmark classifications. In this study, it is used to measure the similarity between detected vulnerabilities and actual vulnerabilities, ranging from 0 to 1, where 1 indicates the highest similarity, suggesting that the predicted values closely match the actual values. The FMI formula is shown in equation (12).

$$FMI = \sqrt{\frac{TP}{TP+FP} * \frac{TP}{TP+FN}} \quad (12)$$

Equations (3) to (12) are commonly used evaluation metrics in binary classification. Accuracy, Precision, and Recall directly evaluate the method from the data, while the F1-Score balances Precision and Recall. The AUC under the ROC curve measures both TPR and FPR, and MCC and FMI are also relatively balanced metrics. Except for FMI, the range of values for all other metrics is between 0 and 1, where higher values indicate better performance.

### B. Experimental Results and Analysis

To validate the effectiveness and efficiency of the proposed method, this section analyzes the experimental results corresponding to the four research questions mentioned previously.

#### 1) Improvement in Fitness Values

Maximizing coverage and accuracy are the two objectives when optimizing static analysis methods for vulnerability detection using a multi-objective optimization algorithm. The algorithm monitors data changes and statement coverage during the static analysis of smart contracts, calculating the improvements in coverage achieved by the multi-objective optimization algorithm while recording changes in detection accuracy. This part of the experiment is based on the 623 manually labeled smart contracts, calculating the average detection results without distinguishing between different types of vulnerabilities.

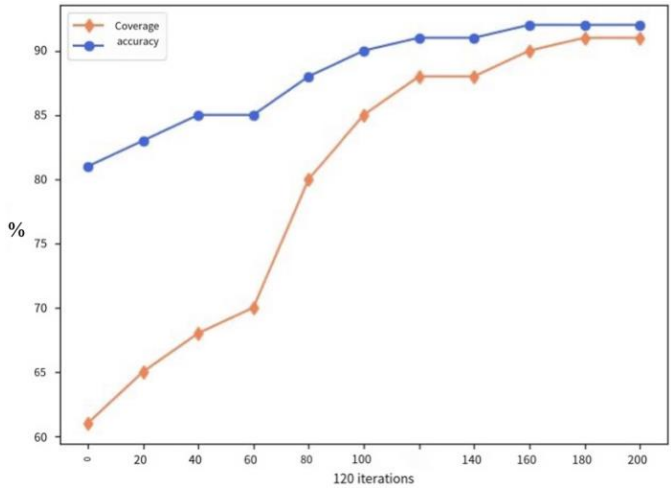

Fig. 4. Coverage and accuracy variation

To assess the impact of the multi-objective optimization algorithm on coverage and accuracy in static analysis methods, all detected values of the objective functions were recorded during the algorithm iterations. The average values of coverage and accuracy were calculated at intervals of 20 iterations. The specific curves for changes in coverage and accuracy are shown in Fig. 4.

As shown in Fig. 4, both coverage and accuracy reach relatively stable values after 120 iterations, with only slight improvements in subsequent iterations. It is also evident that before the multi-objective optimization algorithm was applied (at 0 iterations), the initial statement coverage was relatively low. The multi-objective optimization algorithm significantly improves coverage more than accuracy, as the algorithm enhances data diversity, expanding the coverage of statements during data execution.

The changes in coverage and accuracy were also analyzed for smart contracts of different scales before and after incorporating the multi-objective optimization algorithm into the static analysis method. The specific results are shown in Table VI.

TABLE VI. COVERAGE AND ACCURACY (%)

| Contract Type | Method | Coverage | Accuracy |
|---|---|---|---|
| Simple | Static Analysis | 77.5 | 85.6 |
| | Static Analysis + Multi-Objective | **85.4** | **90.8** |
| Ordinary | Static Analysis | 82.3 | 86.6 |
| | Static Analysis + Multi-Objective | **91.7** | **91.2** |
| Complex | Static Analysis | 87.5 | 87.1 |
| | Static Analysis + Multi-Objective | **97.3** | **93.6** |

As shown in Table VI, simple contracts have a lower coverage rate (only 77.5%) when using static analysis alone, likely due to the smaller amount of data involved, making some statements difficult to cover. For ordinary and complex contracts, which have more lines of code, static analysis covers more statements. After incorporating the multi-objective optimization algorithm to optimize coverage, coverage rates improved for all three contract types. In terms of accuracy, the static analysis method achieves a consistent accuracy of around 86% across all contract types. After incorporating the multi-objective optimization algorithm, accuracy improves to over 90%.

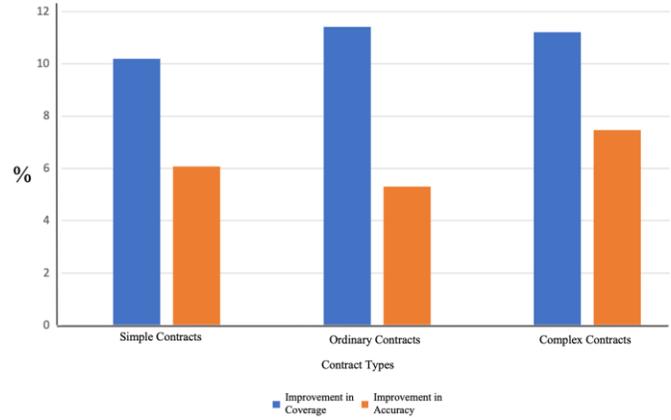

Fig. 5. Statistics of coverage and accuracy improvements

As shown in Fig. 5, the multi-objective optimization algorithm improves coverage by more than 10% for contracts of all three scales. This result indicates that data diversity introduced by the multi-objective optimization algorithm plays an important role in improving statement coverage, with accuracy improvements generally around 6%, with slight fluctuations depending on the contract size.

*2) Efficiency*

The average detection time is also an important metric for evaluating vulnerability detection tools. Current detection tools often take a long time to analyze smart contracts, resulting in low efficiency, especially for tools based on dynamic testing methods. The method proposed in this paper analyzes semantic information using a static analysis approach, which does not require dynamically executing smart contracts, thus reducing the time required to some extent.

TABLE VII. COVERAGE AND ACCURACY (%)

| Vulnerability | Tool | Simple Contracts | Ordinary Contracts | Complex Contracts | Average Time |
|---|---|---|---|---|---|
| Reentrancy | Oyente | 8.5 | 12.6 | 19.3 | 13.4 |
| | Mythril | 24.6 | 35.7 | 45.1 | 35.1 |
| | Proposed Method | **4.2** | **11.3** | **17.4** | **10.9** |
| Call Stack Overflow | Oyente | 11.4 | 12.1 | 22.5 | 15.3 |
| | Mythril | - | - | - | - |
| | Proposed Method | **6.2** | **9.7** | **16.7** | **10.8** |
| Integer Overflow | Oyente | 8.4 | 14.7 | 20.1 | 14.4 |
| | Mythril | 22.1 | 29.7 | 43.2 | 31.6 |
| | Proposed Method | **4.5** | **13.2** | **21.2** | **12.9** |
| Timestamp Dependency | Oyente | 7.3 | 14.8 | 21.9 | 14.6 |
| | Mythril | 26.1 | 31.8 | 42.6 | 33.5 |
| | Proposed Method | **5.2** | **12.5** | **21.2** | **12.9** |

We compared the performance of the proposed method with popular tools such as Oyente [35] and Mythril [36], focusing on the average detection time for various types of vulnerabilities across different contract sizes. Since Mythril cannot detect call stack overflow vulnerabilities, the corresponding data is intentionally omitted. The detailed time statistics are presented in Table VII.

As shown in Table VII, the time cost for detecting vulnerability in simple contracts is relatively low, with Mythril taking the longest time to analyze contracts of the same scale. This is because Mythril uses a more complex method, such as taint analysis and symbolic execution, combining multiple methods to detect vulnerabilities. Oyente, a symbolic execution-based contract analysis tool, simplifies loop statements and uses rule-based pattern matching to detect vulnerabilities, making it slightly faster than Mythril. Table VIII provides statistics on the average detection time for all contract regardless of its scales.

TABLE VIII. AVERAGE DETECTION TIME FOR ALL TARGET CONTRACTS

| Tool | Oyente | Mythril | Proposed Method |
|---|---|---|---|
| Average Time (s) | 14.4 | 33.4 | **11.9** |

Overall, Mythril is the slowest, with an average detection time of 33.4 seconds. In comparison, Oyente and the proposed method are much faster, with Oyente averaging 14.4 seconds, only about 2 seconds longer than the proposed method, which takes just 11.9 seconds on average across simple, ordinary, and complex contracts. Among compared tools, the proposed method demonstrates superior efficiency.

*3) Performance Analysis*

As shown in Fig. 6, the values of TPR, FPR, FNR, and TNR were calculated based on the recorded experimental data for different types of vulnerabilities.

As shown in Fig. 6, the TPR for the four types of vulnerabilities is approximately 85%, while the TNR is around 95%. This is because the manually labeled data contains more negative samples than positive samples, leading to a class imbalance. Additionally, when there are genuinely no vulnerabilities, the likelihood of detecting no vulnerabilities is higher, so most contracts with no detected vulnerabilities indeed have no such vulnerabilities. However, contracts with actual vulnerabilities may still experience some false negatives and false positives after detection. Overall, these results confirm the effectiveness of the proposed method in detecting the four types of vulnerabilities.

To further evaluate the performance of the proposed method, we compared it with state-of-the-art smart contract vulnerability detection tools like Oyente, Mythril, and Slither. These tools can be easily run using Docker, with detection results output directly to JSON files that clearly indicate whether specific vulnerabilities exist or not. We used a manually labeled dataset for vulnerability detection in smart contracts and calculated averages for accuracy, precision, recall, F1-Score, AUC, MCC, and FMI across different vulnerability types. Note that Mythril cannot detect call stack overflows, and Slither cannot detect integer overflow vulnerabilities, so those data rows are intentionally omitted. The specific results are shown in Table IX.

As shown in Table IX, for all four vulnerability types, the proposed method outperforms the others across all seven evaluation metrics. Without considering efficiency, Mythril and Slither perform similarly, while Oyente, which uses simpler methods, yields less favorable results. Since the labeled datasets for the four types of vulnerabilities predominantly contain negative samples, accuracy values are generally high. In terms of precision and recall, the proposed method also outperforms the others, with a particularly noticeable improvement over Oyente, while the improvements over Mythril and Slither are relatively smaller. Overall, the proposed method performs well in detecting these four types of vulnerabilities.

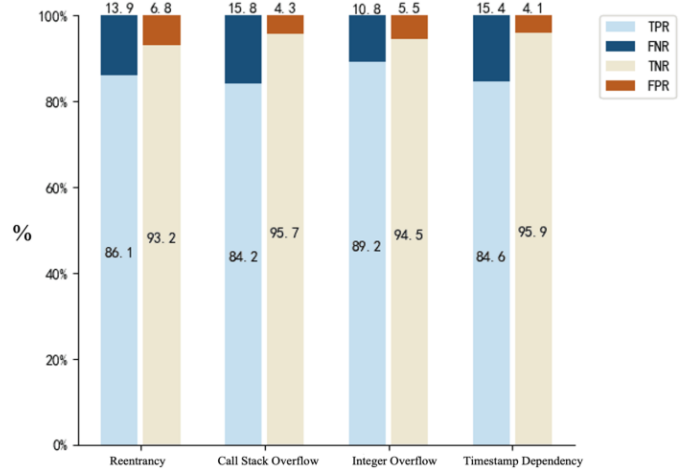

Fig. 6. Distribution of TPR, FPR, FNR, and TNR for various vulnerabilities

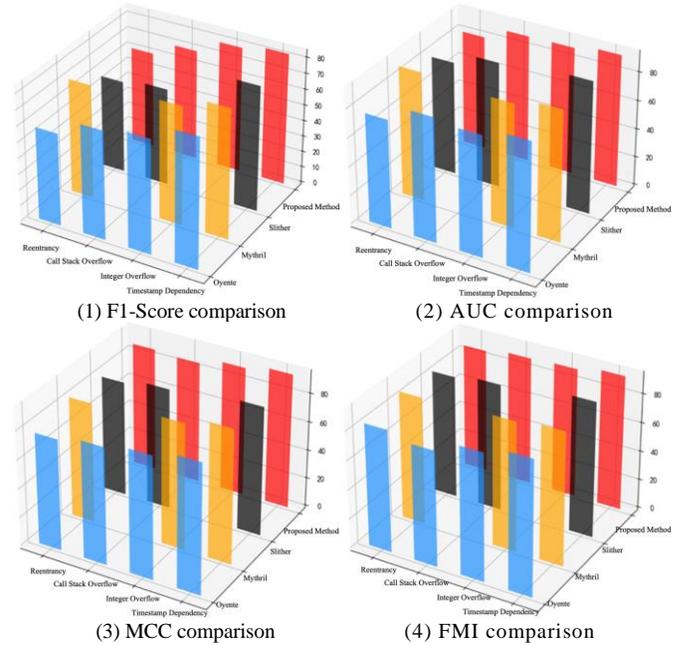

(1) F1-Score comparison  (2) AUC comparison

(3) MCC comparison  (4) FMI comparison

Fig. 7. Visualization of various performance metrics

To further analyze the proposed method, metrics such as F1-Score, AUC, MCC, and FMI were used, and the data were visualized, as shown in Fig. 7. Each column in the figure represents the four vulnerability types—reentrancy, call stack overflow, integer overflow, and timestamp dependency—from left to right, while the rows represent Oyente, Mythril, Slither, and the proposed method, respectively.

From the above metrics, it is evident that the proposed method outperforms Oyente across all four vulnerability types. Mythril and Slither perform comparably, while the proposed method shows a slight improvement over them. The F1-Score of each tool is higher than the AUC, with the AUC improvement being relatively small, but it is comparable to MCC and FMI. Overall, the proposed method slightly outperforms Mythril and Slither in overall classification performance, with a significant improvement over Oyente across all metrics.

To track the changes in F1-Score, AUC, MCC, and FMI values as the number of iterations of the multi-objective optimization algorithm increases, the vulnerability detection results were output at intervals of 20 iterations, and the values of F1-Score, AUC, MCC, and FMI were calculated for each vulnerability type. Both F1-Score and AUC provide insights into the algorithm's effectiveness in detecting actual vulnerabilities, with F1-Score specifically focusing on optimizing the balance between precision and recall, thereby improving the accuracy of the detection process. Besides aiming to increase recall, AUC also aims to reduce the error rate, with higher AUC values indicating lower false positive rates. MCC and FMI consider the overall detection results, providing a more comprehensive and expressive measure. The initial results of vulnerability detection were already obtained using the static analysis method before the algorithm started iterating, so preliminary values were already present at 0 iterations, as shown in Fig. 8.

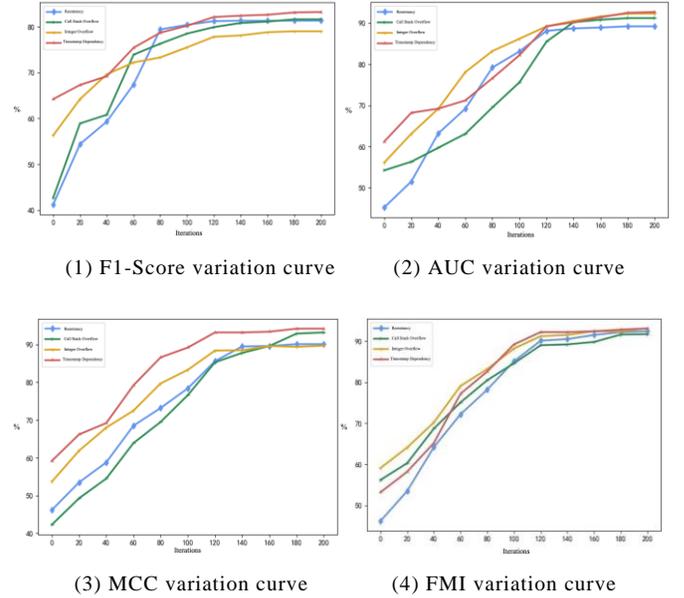

(1) F1-Score variation curve   (2) AUC variation curve

(3) MCC variation curve   (4) FMI variation curve

Fig. 8. Variation curves of metrics with iterations using the proposed method

TABLE IX. COMPARISON OF THE DIFFERENT METHODS ON VARIOUS EVALUATION METRICS (%)

| Vulnerability | Method/Tool | Accuracy | precision | recall | F1-Score | AUC | MCC | FMI |
|---|---|---|---|---|---|---|---|---|
| Reentrancy | Oyente | 81.3 | 78.9 | 56.5 | 58.2 | 75.6 | 77.2 | 83.8 |
|  | Mythril | 85.5 | 86.5 | 67.3 | 68.3 | 86.3 | 81.2 | 79.1 |
|  | Slither | 87.2 | 89.6 | 81.8 | 71.7 | 83.8 | 85.1 | 87.2 |
|  | Proposed Method | **87.9** | **90.1** | **86.1** | **81.3** | **89.2** | **90.1** | **92.4** |
| Call Stack Overflow | Oyente | 92.3 | 83.6 | 63.2 | 71.9 | 89.8 | 83.5 | 87.5 |
|  | Mythril | - | - | - | - | - | - | - |
|  | Slither | 90.2 | 85.3 | 78.6 | 74.8 | 86.3 | 87.9 | 89.1 |
|  | Proposed Method | **94.5** | **89.2** | **84.2** | **81.6** | **91.2** | **93.2** | **91.7** |
| Integer Overflow | Oyente | 89.6 | 71.5 | 49.2 | 58.2 | 79.3 | 80.3 | 84.3 |
|  | Mythril | 91.5 | 69.5 | 54.8 | 61.2 | 87.5 | 84.2 | 87.3 |
|  | Slither | - | - | - | - | - | - | - |
|  | Proposed Method | **92.3** | **92.3** | **89.2** | **79.0** | **92.3** | **89.7** | **93.1** |
| Timestamp Dependency | Oyente | 90.1 | 71.9 | 53.2 | 61.1 | 80.5 | 86.7 | 85.8 |
|  | Mythril | 91.6 | 86.2 | 59.1 | 70.1 | 89.1 | 85.3 | 88.6 |
|  | Slither | 91.3 | 88.4 | 82.6 | 79.6 | 89.4 | 90.3 | 89.5 |
|  | Proposed Method | **93.2** | **89.1** | **84.6** | **83.2** | **92.7** | **94.2** | **93.1** |

As shown in Fig. 8, the growth in each vulnerability type is most significant during the first 120 iterations, after which it gradually stabilizes, with the algorithm having been experimentally proven to converge at around 200 iterations. These four comprehensive metrics are more indicative of the performance of the proposed method compared to the previous metrics, and the proposed method shows good performance across these four metrics for all four vulnerability types, with final values reaching over 90%. Additionally, MCC is mainly used for class imbalance metrics, and the results indicate that the proposed method also performs well in handling class imbalance.

*4) Performance on Unlabeled Data*

To further validate the proposed method's performance on different datasets, experiments were conducted using real smart contracts collected from the Ethereum network that were not manually labeled. The smart contracts were first categorized based on line count, with the numbers of simple, ordinary, and complex contracts already presented earlier. Since the algorithm in this paper involves calculating the accuracy of vulnerability detection, all detected smart contracts must have correct labels assigned beforehand. Therefore, before the experiment, other vulnerability detection tools such as Mythril and Oyente were used to label the smart contracts, and overlapping detection results were considered correct and used as labels. Since Oyente cannot detect call stack overflow vulnerabilities, Securify [37] was used as a supplement for labeling call stack overflow vulnerabilities. The results of labeling 6,070 unlabeled smart contracts using these three tools are shown in Table X.

As shown in Table X, the total number of smart contracts for each of the four vulnerability types is roughly the same, around 1,200-1,700. Categorizing the contracts by scale, the number of simple contracts is relatively small, while the number of ordinary and complex contracts exceeds 2,000 each. The experiment was conducted using these contract scales, recording the values of the two objective functions, coverage and accuracy, during vulnerability detection, and then calculating the average values for all test data, as shown in Table XI.

TABLE X. LABELED SMART CONTRACT CLASSIFICATION AND COUNTS

|  | Reentrancy | Call Stack Overflow | Integer Overflow | Timestamp Dependency | Total |
|---|---|---|---|---|---|
| Simple | 153 | 187 | 201 | 219 | 760 |
| Ordinary | 862 | 514 | 854 | 741 | 2971 |
| Complex | 458 | 526 | 574 | 781 | 2339 |
| Total | 1473 | 1227 | 1629 | 1741 | 6070 |

TABLE XI. COVERAGE AND ACCURACY ON UNLABELED DATA (%)

| Contract Type | Coverage | Accuracy |
|---|---|---|
| Simple | 78.9 | 93.4 |
| Ordinary | 85.3 | 97.2 |
| Complex | 90.2 | 95.8 |

As shown in Table XI, the statement coverage for simple contracts is lower, around 80%, due to the limited number of statements, making full coverage more challenging. Coverage improves as contract complexity increases due to variations in the input data or different execution paths that a smart contract might take during analysis. In terms of accuracy, all contract types achieve high values, likely due to the superior performance of the proposed method compared to the tools used for labeling. As a result, when using data labeled by these tools, the method maintains a high level of accuracy.

TABLE XII. EVALUATION METRICS ON TEST DATA (%)

| Vulnerability | accuracy | precision | recall | F1-Score | AUC |
|---|---|---|---|---|---|
| Reentrancy | 93.2 | 89.2 | 80.6 | 84.6 | 94.8 |
| Call Stack Overflow | 94.3 | 90.3 | 85.7 | 87.9 | 95.8 |
| Integer Overflow | 95.1 | 90.3 | 84.9 | 87.5 | 97.4 |
| Timestamp Dependency | 96.1 | 91.8 | 87.5 | 89.5 | 96.9 |

To evaluate the proposed method's performance on the test data, various metrics were calculated for each vulnerability type, including accuracy, precision, recall, F1-Score, and AUC, as shown in Table XII.

As shown in Table XII, accuracy values are high, largely due to the large number of non-vulnerable cases in the labeled data, which results in a high TN count. Additionally, because the methods used in this paper for labeling selected data are highly accurate, the TP count is also high, further contributing to the high accuracy values of the proposed method. Precision and recall are also strong, indicating reliable detection, and the F1-Score and AUC further confirm the effectiveness of the proposed method. However, since this dataset was labeled using a limited number of tools with limited capability, the reliability of these conclusions is not as strong as it would be with manually labeled datasets used in the previous section.

*C. Threats to Validity*

To ensure effective comparisons, the data used in the experiments were obtained from the literature and the Ethereum network, including both manually labeled and unlabeled data. In the experiments, due to the randomness of the multi-objective optimization algorithm, the results of each vulnerability detection may vary. Therefore, each smart contract's vulnerability detection was performed three times. The results showed that the detection results for different vulnerability types varied little, with randomness mainly affecting coverage changes, likely because the static analysis of smart contracts remained unchanged. The manually labeled data used in this study was derived from manually labeled datasets in peer-reviewed literatures, which are considered reliable, while the unlabeled data was labeled based on detection results from several tools and used to verify the effectiveness of the proposed method. However, the correctness of the tool-labeled datasets does not necessarily represent the actual state of vulnerabilities. Therefore, the evaluation of the experiments in this study may not fully reflect the actual effectiveness of the proposed vulnerability detection method in real-world scenarios, and the results may not be generalizable to other datasets.

V. CONCLUSION

In this paper, we proposed a method that integrates static analysis with the NSGA-II multi-objective optimization algorithm to improve the accuracy and coverage of vulnerability detection in smart contracts. Our approach involved compiling smart contracts into an abstract syntax tree to analyze inheritance, invocation relationships, and data flow, followed by the use of data dependency and variable analysis to detect specific vulnerabilities such as reentrancy, call stack overflow, integer overflow, and timestamp dependencies. By leveraging the NSGA-II algorithm, we optimized detection metrics through a process of multi-objective optimization, simultaneously maximizing accuracy and coverage. Our method was validated using a combination of manually labeled and unlabeled datasets, and demonstrated its superior performance in accuracy, coverage, efficiency, and effectiveness compared to state-of-the-art tools. Future work may focus on integrating dynamic analysis to enhance detection accuracy and reduce false positives and negatives. Additionally, expanding and thoroughly labeling smart contract datasets to better reflect real-world complexities will be essential for further improving and validating the vulnerability detection methods.